\documentclass[epj]{svjour}

\usepackage{epsfig}
\usepackage{amsmath}
\usepackage{amssymb}

\newcommand{\bra}[1]{\mbox{$\langle #1|$}}
\newcommand{\ket}[1]{\mbox{$|#1\rangle$}}

\begin{document}

\title{
\hfill{\small {\bf MKPH-T-05-09}}\\
On the treatment of the $\Delta$-contribution in
 electromagnetic $pp$-knockout reactions}

\author{Carlotta Giusti\inst{1},
        Franco Davide Pacati\inst{1},
 Michael Schwamb\inst{2},
        and 
Sigfrido Boffi\inst{1}}

\institute{
  Dipartimento di Fisica Nucleare e Teorica, Universit\`a degli Studi di Pavia,
 and Istituto Nazionale di Fisica Nucleare, \\
 Sezione di Pavia, I-27100 Pavia, Italy
  \and
  Institut f\"ur Kernphysik, Johannes Gutenberg-Universit\"at
  Mainz, D-55099 Mainz, Germany
}

\date{\today}

\abstract{The treatment of the $\Delta$-current and its contribution in the 
 exclusive $^{16}$O$(e,e'pp)^{14}$C and   $^{16}$O$(\gamma,pp)^{14}$C knockout 
 reactions are investigated in combination with the effects of correlations. 
 Different parametrizations of the effective $\Delta$-current   
 and different treatments of correlations in the two-nucleon overlap function
 are considered. 
 The results are presented and discussed for a suitable 
 choice of kinematics. It is found that the investigation of
 different mutually supplementing 
 kinematics is necessary to resolve the uncertainties in the
 theoretical ingredients and extract clear and unambiguous information on
 correlations.}


\titlerunning{ }

\authorrunning{ }

\date{\today}

\maketitle

\section{Introduction}
\label{sec1}

The independent particle shell model, describing a nucleus as a system
 of nucleons moving in a mean field, can reproduce many basic features of 
 nuclear structure. It is, however, nowadays understood that the repulsive
core of the $NN$-interaction induces additional short-range
 correlations (SRC) which are beyond a mean-field description.  SRC 
 have a decisive influence on the spectral distribution of nucleons and 
 on  the binding properties of atomic nuclei. A powerful tool for the study
 of SRC are  electromagnetically induced two-nucleon knockout reactions
 like $(\gamma,NN)$ or $(e,e' NN)$ because the probability 
 that a real or a virtual photon is absorbed by a nucleon pair  should be
 a direct measure for SRC (for an overview, see \cite{BoG96}).
 However, this simple picture is modified by the competing mechanisms 
 which may additionally contribute to two-nucleon knockout due to their
 two-body character. 
 At low and intermediate energies, the most important ones are those due to 
 two-body meson-exchange (MEC) and $\Delta$ currents, 
 as well  as final-state interactions (FSI).
  The latter consist in principle of two different contributions, namely the 
 mutual interaction of the two emitted nucleons ($NN$-FSI), which can be 
 described by a realistic $NN$-interaction \cite{ScB03,ScB04}, and the 
 interaction of each of the two outgoing nucleons  with the residual nucleus, 
 which is  described in our model by a suitable optical  potential.  
 Whereas $NN$-FSI depend strongly on the kinematics and on the chosen 
electromagnetic probe \cite{ScB03,ScB04}, the $N$-nucleus interaction generally 
represents the main contribution of FSI and can never be neglected. The optical
potential always leads to a strong reduction of the calculated cross 
 section leaving, however, its main qualitative features 
 unchanged  \cite{ScB03,ScB04,BGP}. Moreover, the model dependence
 due to the specific  choice of a realistic optical potential turns out to be
 small  \cite{BoG96,BGP}.  

\begin{figure}[h]
\centerline{
\resizebox{0.5\textwidth}{!}{
  \includegraphics{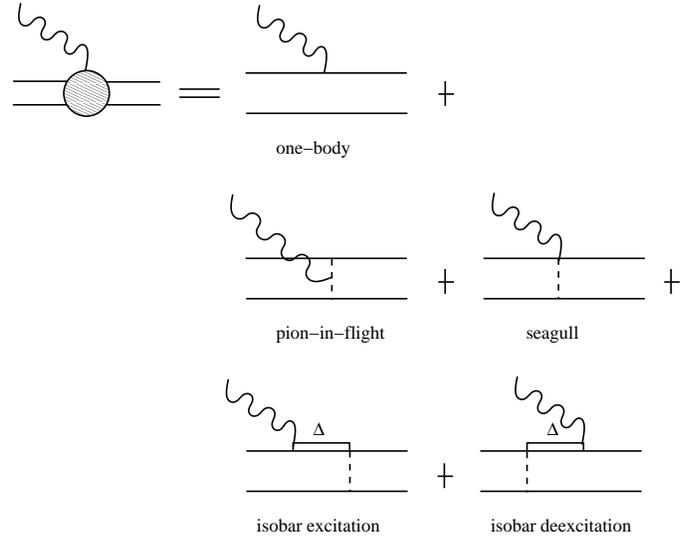}
}}
\caption{The electromagnetic currents contributing to two-nucleon knockout
 reactions at low and intermediate energies.}
\label{fig01}
\end{figure}

 Concerning the two-body currents, the nonrelativistic pion-in-flight and 
 seagull MEC contri\-butions  (see Fig. \ref{fig01}) are forbidden in 
 $pp$-knockout. Therefore, the only electromagnetic background mechanism we 
 have to deal with in  $pp$-knockout is the $\Delta$-current, 
 consisting of an excitation and a deexcitation part (Figure \ref{fig01}).

 It was found in previous studies
 \cite{BoG96,GiP97,GPA98,GiP98,Ryc94,Ryc97,Ryc04,Co03,Co04} that the relevance 
 of the $\Delta$-contribution depends strongly on the kinematics 
 and on the particular 
 final state of the residual nucleus, as well as on the type of electromagnetic 
 probe. It is possible to envisage specific situations where either 
 the contribution
 of the one-body or of the two-body $\Delta$-current is dominant. A combined
 study of both types of situations may provide an interesting tool to
 disentangle and separately investigate the two reaction processes.  
 
 If one is primarily interested in studying  correlations, 
 situations and kinematics should be preferred where the 
 $\Delta$-contribution  
 is as small as possible, because the remaining one-body contribution can only 
 contribute via correlations in the initial or in the final state, and should 
 therefore be maximized in order to have the most direct access to  SRC.
 
 In any case, a conceptual satisfying description of the behaviour of the 
 $\Delta$ in the nuclear medium is necessary.
 This  would also require a consistent  treatment of nucleonic as well as
 $\Delta$-degrees of freedom in the two-body overlap function in the initial
 as well as in the final state which is, however, not available at present.
 Therefore, one has to rely on approximative  schemes. 

 The central aim of the present paper is a systematic study  of different 
 tractable parametrizations of the $\Delta$-current and of their contribution 
 to $pp$-knockout off complex nuclei in comparison with the one-body one. 
 The paper is organized as follows. In sect. \ref{sec2} various 
 treatments of the $\Delta$-current are discussed. In sect. \ref{sec3} the 
 choice of kinematics for the present investigation is discussed. Numerical 
 results for the cross sections of the exclusive $^{16}$O$(e,e'pp)^{14}$C 
 and $^{16}$O$(\gamma,pp)^{14}$C reactions are presented  
 in sect. \ref{sec4}. Some conclusions are drawn in sect. 5. 
 
\section{The $\Delta$-current}\label{sec2}
As has already been mentioned, the effective $\Delta$-current operator 
$\vec{\jmath}_{\Delta N}$, 
depicted in the bottom line of Fig.\ \ref{fig01}, consists of two parts, 
 namely an excitation $(I)$ and a deexcitation $(II)$ part
\begin{equation}\label{delta1}
\vec{\jmath}_{\Delta N } =   \vec{\jmath}^{(I)}_{\Delta N }(1,2) +
 \vec{\jmath}^{(II)}_{\Delta N }(1,2) + \left( 1 \rightarrow 2 \right)\,\, ,
\end{equation}   
which are given by
\begin{eqnarray}\label{delta2}
\vec{\jmath}^{(I)}_{\Delta N }(1,2) &=&  V_{N \Delta}(1,2) 
G_{\Delta}(\sqrt{s_I})
 \vec{{\cal J}}_{\Delta N}(1)\,\, , \label{delta2_1}\\
\vec{\jmath}^{(II)}_{\Delta N } &=&  
 \vec{{\cal J}}_{N \Delta}(1) G_{\Delta}(\sqrt{s_{II}})
V_{\Delta N}(1,2) \,\, . \label{delta2_2}
\end{eqnarray}
In these expressions  $\vec{{\cal J}}_{\Delta N}$,
with $\vec{{\cal J}}_{\Delta N}$=$(\vec{{\cal J}}_{N \Delta})^{\dagger}$,
 describes the electromagnetic
 transition $\gamma N \rightarrow \Delta$. If we restrict ourselves to the
 dominant magnetic dipole (M1) transition, $\vec{{\cal J}}_{\Delta N}$
 is given in photonuclear reactions by the matrix element 
\begin{equation}\label{delta3}
\bra{\Delta} \vec{{\cal J}}_{\Delta N} \ket{N} =
 \frac{G^{\Delta N}_{M1}}{2 M_N} i \vec{\sigma}_{\Delta N} \times \vec{k}
 \,\, e \left( \tau_{\Delta N} \right)_0\,\, , 
\end{equation}
where $e$ denotes the elementary charge, $\vec{k}$ the photon
 momentum  and $M_N$ the nucleon mass.  The spin and isospin transition
 operators $\vec{\sigma}_{\Delta N}$ and $\vec{\tau}_{\Delta N}$
 are fixed by their reduced matrix elements 
 $\bra{\frac{3}{2}}|{\sigma}_{\Delta N} | \ket{\frac{1}{2}} = 2$ and
 $\bra{\frac{3}{2}}|{\tau}_{\Delta N} | \ket{\frac{1}{2}} = 2$.
The value of the coupling constant  $G^{\Delta N}_{M1} = 4.22$ can be 
extracted from the elementary total photopionproduction cross section in the 
$\Delta$-region \cite{Sch95}.  For virtual photons a usual electromagnetic
 dipole form factor 
\begin{equation}\label{dipol1}
F(Q^2)= \left[ 1 + \frac{Q^2}{(855 \mbox{MeV})^2}\right]^{-2}
\end{equation}
has additionally to be taken into account. 

The propagator $G_{\Delta}$ in (\ref{delta2}) depends strongly on the
 invariant energy $\sqrt{s}$ of the $\Delta$. If we omit 
 medium modifications and treat the $\Delta$ as a free particle,
 we use, following \cite{WiA97},
\begin{equation}\label{prop1}
G_{\Delta}(\sqrt{s}) = \frac{1}{M_{\Delta} - \sqrt{s} - \frac{i}{2}
 \Gamma_{\Delta}(\sqrt{s})}\,\, , 
\end{equation}
 where $\Gamma_{\Delta}$ is the energy-dependent decay width of the
$\Delta$ taken from Ref. \cite{BM73} and $M_{\Delta}=1232$ MeV its mass. 
 In the excitation part, we use for the invariant energy 
 $\sqrt{s_I} = \sqrt{s_{NN}} - M_N$, where $\sqrt{s_{NN}}$ is the 
 experimentally measured invariant energy of the two outgoing protons.
 In the  deexcitation part the choice   $\sqrt{s_{II}} = M_N$
 turns out to be the most appropriate one \cite{WiA97}.

We now turn to the potential  $V_{N \Delta}(1,2)$, 
with $V_{\Delta N}(1,2)$ = $\left(V_{N \Delta}(1,2)\right)^{\dagger}$, 
describing the transition $ N \Delta \rightarrow NN$ via meson exchange.
In this work, besides the usual static $\pi$-exchange, we consider in addition 
also  the static $\rho$-exchange, i.e.
\begin{equation}\label{ndpot1} 
V_{N \Delta} = V^{\pi}_{N \Delta} +  V^{\rho}_{N \Delta}\,\, , 
\end{equation}
whose explicit expressions are well known from literature, see e.g. 
\cite{Mac89,GaM90,VanM}:
\begin{eqnarray}
\bra{NN(\vec{p}^{\,\prime})}  V^{\pi}_{N \Delta} \ket{\Delta N (\vec{p}\,)}
&=& -\frac{1}{\left( 2 \pi \right)^3} \, F_{\pi NN}(q^2) \, 
F_{\pi N \Delta}(q^2)\, \nonumber \\
 & & 
\!\!\!\!\!\!\!\!\!\!\!\!\!\!\!\!\!\!\!\!\!\!\!
\!\!\!\!\!\!\!\!\!\!\!\!\!\!\!\!\!\!\!\!\!\!\!
\!\!\!\!\!\!\!\!\!\!\!\!\!\!\!\!\!\!\!\!\!\!\!
\frac{f_{\pi NN} f_{\pi N \Delta}}{m_{\pi}^2} 
\vec{\tau}_{NN}(2) \cdot \vec{\tau}_{N \Delta}(1)  
\frac{\vec{\sigma}_{NN}(2) \cdot \vec{q} \, 
 \vec{\sigma}_{N\Delta}(1) \cdot \vec{q}}{q^2 + m_{\pi}^2}\,\, ,
\label{ndpotpi1}
\end{eqnarray}
\begin{eqnarray}
\bra{NN(\vec{p}^{\,\prime})}  V^{\rho}_{N \Delta} \ket{\Delta N (\vec{p}\,)}
&=& -\frac{1}{\left( 2 \pi \right)^3} \, F_{\rho NN}(q^2) \, 
F_{\rho N \Delta}(q^2)\, \nonumber \\
 & & 
\!\!\!\!\!\!\!\!\!\!\!\!\!\!\!\!\!\!\!\!\!\!\!\!
\!\!\!\!\!\!\!\!\!\!\!\!\!\!\!\!\!\!\!\!\!\!\!\!
\!\!\!\!\!\!\!\!\!\!\!\!\!\!\!\!\!\!\!\!\!\!\!\!
\frac{f_{\rho NN} f_{\rho N \Delta}}{m_{\rho}^2} 
\vec{\tau}_{NN}(2) \cdot \vec{\tau}_{N \Delta}(1)  
\frac{\left( \vec{\sigma}_{NN}(2) \times \vec{q}\right)   \cdot  
 \left(\vec{\sigma}_{N\Delta}(1) \times  \vec{q}\right) }{q^2 + m_{\rho}^2}
\,\, . \nonumber \\
\label{ndpotrho1}
\end{eqnarray}
 In these expressions  $\vec{q} = \vec{p} - \vec{p}^{\prime}$, where 
 $\vec{p}$ ($\vec{p}^{\prime}\,$) denotes the relative momentum of the
 $\Delta N$ ($NN$) system in the initial (final)  state. 
  The quantities $F_{x NN}$ and   
$F_{x N \Delta}$, $x \in \{\pi, \rho \}$, are the so called hadronic
 form factors necessary for regularizing the potentials at short distances,
 where the meson-exchange picture becomes meaningless. As usual, they are 
 parametrized as follows
\begin{eqnarray}
 F_{x NN}(q^2) &=& \left(\frac{ \Lambda^2_{x NN} - m^2_{x}}{\Lambda_{xNN}^2 +
 q^2 }\right)^{n_{x NN}}\,\, , \label{form1} \\
 F_{x N \Delta}(q^2)  &=& \left(\frac{ \Lambda^2_{x N\Delta} - 
m^2_{x}}{\Lambda_{xN\Delta}^2 +
 q^2 }\right)^{n_{x N\Delta}}\,\, , \label{form2}
\end{eqnarray}
 where the integers $n_{x NN}$, $n_{x N\Delta}$, as well as the cutoffs
 $\Lambda_{x NN}$ and $\Lambda_{x N\Delta}$, can be treated as free parameters.
 In the following, various approaches are considered where 
 different values are given to these parameters, as well as to the coupling 
 constants $f_{x NN}$ and $f_{x N \Delta}$.

 In the simplest approach, called $\Delta$(NoReg), we use only an 
 unregularized pionic transition potential, i.e.
\begin{eqnarray}
& & 
\frac{f^2_{\pi NN}}{4 \pi} = 0.08\, , 
\frac{f^2_{\pi N \Delta}}{4 \pi} = 0.35\, , 
\Lambda_{\pi NN} = \Lambda_{\pi N \Delta} \rightarrow \infty, \nonumber \\
& &  V^{\rho}_{N \Delta} =0 \,\, .
\end{eqnarray}
 Here, $f_{\pi N \Delta}$ has been  extracted from the $\Delta$-decay width.  
An unregularized pionic transition potential is used in 
\cite{Co03,Co04,Co93}. This approach is similar to our 
previous treatment of the $\Delta$-current, where only a simple 
regularization was included in  coordinate space. 

\begin{figure}[h]
\centerline{
\resizebox{0.5\textwidth}{!}{
  \includegraphics{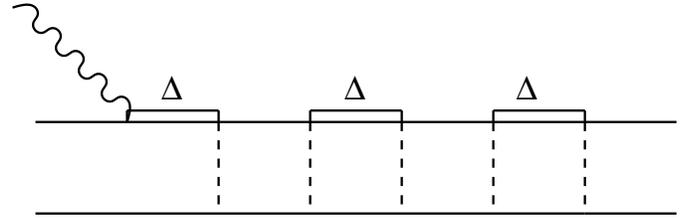}
}}
\caption{A possible $\Delta$-contribution to $pp$-knockout which is of higher
 order in $f_{x N \Delta}$ and leads to divergences within an 
 unregularized approach for $V_{N \Delta}$.}
\label{fig02}
\end{figure}

An unregularized transition potential can be used only if the  
$\Delta$-contribution to $pp$-knockout is performed perturbatively 
 up to the first order in $f_{\pi N \Delta}$. 
 In a more sophisticated approach going beyond (\ref{delta2}), 
 additional contributions like the one  depicted in 
 Fig.\ \ref{fig02} could in principle occur,  where the $\Delta$ can 
 be excited and deexcited several times 
in the  nuclear medium.  Such mechanisms lead to serious  
 divergences, well known from $NN$-scattering \cite{Mac89,MaH87},
which can only  be removed within a regularized treatment.
 The essential question is, however, how to fix, within such a refined  
 approach, the free parameters, especially the cutoffs,  
 in (\ref{ndpotpi1}) and (\ref{ndpotrho1}). In this work, we select
 for this purpose two alternative scenarios: {\it $NN$}-scattering and 
 {\it $\pi N$}-scattering.

In a first approach, that we call $\Delta(NN)$, the parameters are 
fixed considering the {\it NN}-scattering in the $\Delta$-region.
  For this purpose, we use a nonperturbative treatment of $V_{N\Delta}$ as 
  outlined in some detail in \cite{Wil92,ScA01}. It turns out that a fairly 
 good  description of the $NN$-scattering data in the $\Delta$-region can be 
 achieved by choosing  parameters similar to the ones of the full Bonn 
 potential \cite{MaH87}, i.e.
\begin{eqnarray}
\frac{f^2_{\pi NN}}{ 4\pi} =0.078, \, & &
\frac{f^2_{\pi N \Delta}}{ 4\pi} =0.224, \, \nonumber \\
\frac{f^2_{\rho NN}}{ 4\pi} =7.10, \, & &
\frac{f^2_{\rho N\Delta}}{ 4\pi} =20.45, \, \nonumber \\
\Lambda_{\pi NN} = 1300 \,\mbox{MeV},\, & &
\Lambda_{\pi N \Delta} = 1200 \,\mbox{MeV},\, \nonumber \\
\Lambda_{\rho NN} = 1400 \, \mbox{MeV},\, & &
\Lambda_{\rho N \Delta} = \, 1000\, \mbox{MeV},\, \nonumber \\
& & 
\!\!\!\!\!\!\!\!\!\!\!\!\!\!\!\!\!\!\!\!\!\!\!\!\!\!\!\!\!\!\!
\!\!\!\!\!\!\!\!\!\!\!\!\!\!\!\!\!\!\!\!\!\!\!\!\!\!\!\!\!\!\!
n_{\pi NN} = n_{\pi N\Delta} = n_{\rho NN} =n_{\rho N\Delta} = 1\,\, .
\label{para3}
\end{eqnarray}

In an alternative approach, the parameters of the $\Delta$ are fixed from 
{\it $\pi N$}-scattering in the $P_{33}$-channel.
 We call this approach {\bf $\Delta(\pi N)$}. If we 
consider only the dominant $P_{33}$-channel, a suitable choice of parameters 
is \cite{PoS87}
\begin{eqnarray}
\frac{f^2_{\pi N \Delta}}{ 4\pi} =1.393, \, & & 
\Lambda_{\pi N \Delta} = 287.9 \, \mbox{MeV}, n_{\pi N \Delta} =1\,\,.
\label{para21}
\end{eqnarray}
 The values for $f_{\pi N N}$ and $\Lambda_{\pi NN}$ cannot be simply extracted
 from $\pi N$-scattering in the $P_{33}$-channel.
 Therefore, here we use the ones of the full Bonn 
 potential \cite{MaH87}, i.e. the same as in  (\ref{para3}):
\begin{eqnarray}
\frac{f^2_{\pi N N}}{ 4\pi} =0.078, \, & & 
\Lambda_{\pi N N} = 1300\, \mbox{MeV}, n_{\pi N N} =1\,\,.
\label{para22}
\end{eqnarray}
We note that also in this approach, like in $\Delta$(NoReg),  the 
$\rho$-exchange part $V^{\rho}_{N \Delta }$ is switched off.
We point out the dramatic differences in the values of $f_{\pi N\Delta}$ and 
$\Lambda_{\pi N \Delta}$  for the two treatments in (\ref{para3})  
 and (\ref{para21}).
 A more detailed discussion of this point can be found in 
\cite{TaO85,ElH88,HoM97}. 

In all the approaches considered till now, the $\Delta$ is treated as a free 
particle. In $pp$-knockout on complex nuclei, however, medium modifications in 
the $\Delta$-excitation or deexcitation mechanism  may occur. In order to 
obtain an indication of the relevance  of medium effects, we follow the 
procedure suggested in \cite{ChL88} by a comparison between the 
$^{12}$C$(e,e')$ cross section in the $\Delta$-region 
 and the results of  the $\Delta$-hole model \cite{KoM84,KoO85}, and add a 
 shift $(-30 - 40i)$ MeV in the denominator of $G_{\Delta}$  (see 
 eq. (\ref{prop1})).  
 In this last approach, that we call $\Delta (\pi N, \mathrm{mod})$, we use 
 such a modified $G_{\Delta}$  propagator and the same coupling constants and 
 cutoffs as in $\Delta(\pi N)$. A similar approach was used in the analysis 
 of \cite{McG}.

\section{The choice of kinematics}\label{sec3}

   A consistent treatment of the $\Delta$ in electromagnetic 
 break\-up reactions on  complex nuclei is presently not available.
 The parametrizations of the effective $\Delta$-current proposed in the 
 previous section are therefore not ``fundamental''  from a certain point of 
 view. By choosing extreme scenarios, like an unregularized versus strongly 
 regularized treatments, we are able to obtain an estimate of the 
 theoretical uncertainties. Some explicit results are presented in
 the next section. It is however clear from the beginning that these
 different parametrizations may give large numerical differences in the 
 observables. From this point of view, one might be interested
 in specific kinematics where the $\Delta$-contribution as well as its
 sensitivity to the different parametrizations is either maximized (i)
 or minimized (ii). Case (i) allows us to pin down  the 
 most suitable $\Delta$-parametrization  in two-nucleon knockout
 whereas case  (ii) represents in principle the cleanest scenario to extract 
 correlation effects.

Specific kinematics where either contribution of the one-body or of the 
two-body $\Delta$-current is dominant were already envisaged in previous
studies of electromagnetic two-nucleon knockout. These kinematics can represent
a good basis for the present investigation. Since, however, our main aim here is
to evaluate the relevance of the unavoidable uncertainties in the 
$\Delta$-contribution, it can be useful to maximize or minimize these 
uncertainties with the help of interference effects between the $\Delta$- and 
the one-body current. We may try to study this problem analytically using some  simplifying assumptions. Therefore, let us ignore for the moment FSI, so that 
 the final state of the two outgoing nucleons can be described by an 
 antisymmetrized (${\cal A}$) plane wave (PW):
\begin{eqnarray}
\ket{\psi_f}_{\cal A} &=& 
\ket{\vec{p}_1, \vec{p}_2;s_f,m_{s_f}; t_f,m_{t_f}}\nonumber \\
 & & - (-1)^{s_f+t_f}
\ket{\vec{p}_2, \vec{p}_1;s_f,m_{s_f}; t_f,m_{t_f}}\,\, . 
\label{final1}
\end{eqnarray}
It is known from previous investigations that such a simple PW-approximation  
leaves in most kinematics the qualitative features of the cross sections 
unchanged.
 In (\ref{final1}), $\vec{p}_i$ denotes the asymptotic free momentum of nucleon $i$, while
 $s_f$ and $m_{s_f}$  label the total spin  of the two outgoing nucleons and 
 its projection, respectively. For the isospin quantum numbers $t_f$ and  
 $m_{t_f}$, in  $pp$-knockout we have always $t_f=1$ and $m_{t_f}=1$.

In the $(e,e'pp)$ reaction the one-body current consists of three parts, 
namely the longitudinal charge term ($\rho$) and the transverse convection 
($\vec{\jmath}^{con}$) and spin ($\vec{\jmath}^{spin}$) currents.
 Denoting by $\ket{\psi_i}$ the relative
 part of the initial state with spin (isospin)
 quantum numbers $s_i, m_{s_i}$ ($t_i, m_{t_i}$) and
 wave function $\psi_i$, the relevant
 matrix elements of the three terms of the one-body current are given by  
\begin{eqnarray}
_{\cal A}\bra{\psi_f} \rho(1) \ket{\psi_i} &\sim&
\delta_{s_f s_i} \delta_{m_{s_f} m_{s_i}} \nonumber \\ & &
\!\!\!\!\!\!\!\!\!\!\!\!\!\!\!\!\!\!\!\!\!\!
\!\!\!\!\!\!\!\!\!\!\!\!\!\!\!\!\!\!\!\!\!\!
\times \,  \left[ \psi_i\left(\vec{p}-\frac{\vec{k}}{2}\right) +(-1)^{s_f}
 \psi_i\left(-\vec{p}-\frac{\vec{k}}{2}\right) \right] \,\, , \label{charge1}
\\
_{\cal A}\bra{\psi_f} \vec{\jmath}^{con}(1) \ket{\psi_i} &\sim&
\left( \frac{2 \vec{p}_1  -\vec{k}}{2 M_N} \right)
 \delta_{s_f s_i} \delta_{m_{s_f} m_{s_i}} \nonumber \\ & &
\!\!\!\!\!\!\!\!\!\!\!\!\!\!\!\!\!\!\!\!\!\!
\!\!\!\!\!\!\!\!\!\!\!\!\!\!\!\!\!\!\!\!\!\!
\times \,  \left[ \psi_i\left(\vec{p}-\frac{\vec{k}}{2}\right) +(-1)^{s_f}
 \psi_i\left(-\vec{p}-\frac{\vec{k}}{2}\right) \right] \,\, , \label{con1}
\\
_{\cal A}\bra{\psi_f} \vec{\jmath}^{spin}(1) \ket{\psi_i} &\sim&
 \bra{s_f m_{s_f}} i \frac{\vec{\sigma}_{NN}(1) \times 
 \vec{k}}{2 M_N} \ket{s_i m_{s_i}}  \nonumber \\ & &
\!\!\!\!\!\!\!\!\!\!\!\!\!\!\!\!\!\!\!\!\!\!
\!\!\!\!\!\!\!\!\!\!\!\!\!\!\!\!\!\!\!\!\!\!
 \times \, \left[ \psi_i\left(\vec{p}-\frac{\vec{k}}{2}\right) +(-1)^{s_f}
 \psi_i\left(-\vec{p}-\frac{\vec{k}}{2}\right)  \right]\,\, , \label{spin1}
\end{eqnarray}
where isospin factors have been omitted for the sake of simplicity.
 The quantity $\vec{p}$ denotes the relative momentum of the two
 nucleons in the final state, i.e. 
 $\vec{p} = \frac{ \vec{p}_1 - \vec{p}_2}{2}$. 

\begin{figure*}[htp]
\centerline{
\resizebox{0.2\textwidth}{!}{
  \includegraphics{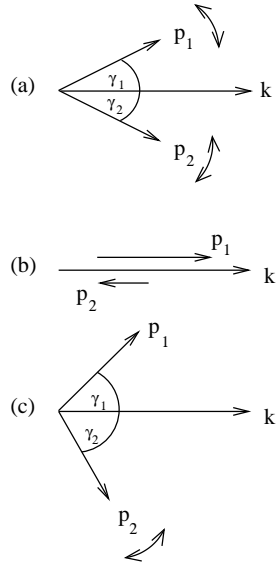}
}}
\caption{
Graphical illustration of the selected kinematics: (a) symmetrical
 kinematics with $|\vec{p}_1|$ = $|\vec{p}_2|$, $|\gamma_1|$=
 $|\gamma_2|$; (b) superparallel kinematics; (c) kinematics with
 $\gamma_1=45^{\circ}$ fixed and $\gamma_2$ varying on the other side
 of the photon momentum $\vec{k}$.}
\label{fig03}
\end{figure*}

It is obvious from the above equations that the spin-term  (\ref{spin1}) 
 is the ideal interference partner for the $\Delta$-current, because both
  have their main contribution in  the magnetic dipole ($M1$)-transition. 
 When the two protons are in a $^1S_0$ initial relative state, $s_i=0$ and  
$s_f$ must be  necessarily 1 for the spin-current contribution (\ref{spin1}). 
 As a consequence,  if 
 $|\vec{p} - \frac{\vec{k}}{2}|$ =   $|-\vec{p} - \frac{\vec{k}}{2}|$,
 the spin-current vanishes and cannot produce any interference with the 
 $\Delta$-current.  This is just the condition of the so-called symmetrical 
 kinematics, depicted in the top panel of Fig. \ref{fig03}, where the two
 nucleons are ejected at equal energies and equal but opposite angles with
 respect to the momentum transfer. As a sidemark, we would like to mention that
 in this kinematics, besides the spin-current, also the 
 convection part (\ref{con1}) 
 vanishes within a PW-approach for a $^1S_0$ initial relative state.

 In contrast, in the so-called
 super-parallel  kinematics, depicted in the middle panel of Fig. \ref{fig03},
 where the two nucleons are ejected parallel and anti-parallel to the momentum
 transfer $\vec{k}$, the difference of the arguments 
$|\vec{p} - \frac{\vec{k}}{2}|$ and $|-\vec{p} - \frac{\vec{k}}{2}|$ 
 of $\psi_i$ in (\ref{spin1}) is maximized  for a given $\vec{k}$,
 and the contribution of the spin-current should become important.  
 Therefore, we can expect that in the super-parallel kinematics also 
 the interference between the spin- and  the $\Delta$-contribution 
 is maximized and the existing uncertainties in the 
 $\Delta$-parametrization become most crucial.
 
 These arguments, which have been brought within a PW-approach and for a 
 $^1S_0$ initial proton pair, should not be changed significantly by 
 FSI and should be valid in all the situations where the $^1S_0$ relative 
 partial wave gives the main contribution, such as in the 
 $^{16}$O$(e,e'pp)$ reaction to the $0^+$ ground state of $^{14}$C \cite{GPA98}.  
 Therefore, in the present investigation cross section calculations for the 
reaction $^{16}$O$(e,e'pp)^{14}$C$_{\mathrm{g.s.}}$  have been performed in  
coplanar symmetrical and super-parallel kinematics, i.e. in two situations 
where the contribution of  the $^1S_0$ relative wave, and therefore also of 
SRC, is emphasized  \cite{GPA98}, and where the interference effects between 
the one-body and the $\Delta$-current should be minimized (symmetrical 
kinematics) and maximized (super-parallel kinematics). 

The super-parallel kinematics chosen for the calculations is the same already 
considered in our previous work \cite{ScB03,ScB04,GPA98,BGPD04} and realized in
the $^{16}$O$(e,e'pp)^{14}$C experiment at MAMI~\cite{Rosner}. The incident
electron energy is $E_{0}=855$ MeV, the energy transfer $\omega=215$ MeV, 
and $k=316$ MeV/$c$. Different values of the recoil momentum 
$\vec{p_{\mathrm{B}}} = \vec{k} - \vec{p_1} - \vec{p_2}$ are obtained 
changing the kinetic energies of the two outgoing protons. 
The symmetrical kinematics is calculated with the same values of  $E_{0}$,  
$\omega$, and $k$, the kinetic energies of the two outoing nucleons are 
determined by energy conservation and different values of  $p_{\mathrm{B}}$ 
are obtained changing the scattering angles of the two protons.

Two different kinematics are considered also for the reaction 
$^{16}$O$(\gamma,pp) ^{14}$C$_{\mathrm{g.s.}}$: a coplanar symmetrical 
kinematics with an incident photon energy $E_\gamma=400$ MeV,  and the coplanar 
kinematics depicted in the bottom panel of  Fig. \ref{fig03} with 
$E_\gamma=120$ MeV, where the energy and the scattering angle of the first 
outgoing proton are fixed, at $T_1= 45$ MeV and $\gamma_1= 45^{\circ}$, 
respectively, the kinetic energy of
the second proton is determined by energy conservation, and different values of 
$p_{\mathrm{B}}$  are obtained by varying the scattering angle
 $\gamma_2$ of the second outgoing nucleon on the other side of the
 photon momentum.  This choice of kinematics for the $(\gamma,pp)$ reaction
 is determined by the results of our previous work \cite{ScB04,GiP98}. 
 It was found in  \cite{GiP98} that the symmetrical kinematics is dominated by 
 the $\Delta$-current, whereas the contribution of the one-body spin- and 
 convection-currents is only of minor importance. Therefore, this case is 
 interesting to give direct access to the 
 $\Delta$-contribution in a situation where interference effects between the 
 one-body and the $\Delta$-current are expected to be small.
 Note the difference in the symmetrical kinematics for $(\gamma,pp)$  and 
 $(e,e'pp)$: As discussed above, in both cases 
the contributions of the transverse convection- and 
 spin-current are  suppressed, but in $(e,e'pp)$ also the longitudinal 
 charge-term contributes and is dominant, as will become apparent in the next 
 section. On the other hand, in the kinematics at 
$E_\gamma=120$ MeV both one-body and $\Delta$-current 
contributions are important \cite{ScB04}. Therefore, this case can be helpful 
to investigate the interplay between one-body and two-body currents in 
the $(\gamma,pp)$ reaction.

\section{Results}\label{sec4}

The cross section of the exclusive  $^{16}$O$(e,e'pp) ^{14}$C and \\
$^{16}$O$(\gamma,pp) ^{14}$C reactions have been calculated in the kinematics
discussed in the previous section. The theoretical model is the same already 
presented in \cite{GiP97,GPA98,BGPD04}.

The basic ingredients of the calculation are the matrix elements of the 
nuclear charge-current operator between initial and final nuclear
many-body   states, i.e.,
\begin{equation}\label{eq1}
J^{\mu}(\vec{k}\,) = \int \bra{\Psi_f} j^{\mu}(\vec{r}) \ket{\Psi_i}
e^{i \vec{k} \cdot \vec{r}} {\mathrm d}\vec{r}\,\, .
\end{equation}
Bilinear products of these integrals give the components of the 
hadron tensor, whose suitable combinations give all the observables available 
from the reaction process \cite{BoG96}. 

The model is based on the two assumptions of an exclusive knockout reaction: 
the direct mechanism and the 
transition to a specific discrete state of the residual nucleus
\cite{GiP97,GPA98,GiP91}. Thus, we consider a 
direct one-step process where the electromagnetic probe directly interacts 
with the  pair of nucleons that are emitted and the A$-$2 $\equiv$ B 
nucleons of the residual nucleus behave  as spectators. Recent experiments 
\cite{Rosner,Ond97,Ond98,Sta00,McG95,Lam96,McG98,Wat00} on reactions induced
by real and virtual photons have confirmed the validity of this mechanism for 
low values of the excitation energy of the residual nucleus. 

As a result of these two assumptions, the integrals (\ref{eq1}) can be reduced 
to a form with three main ingredients: the two-nucleon overlap function (TOF) 
$\ket{\psi_i}$ between the ground state of the target and the final state of the residual 
nucleus, the nuclear current ${\jmath}^{\mu}$ of the two emitted 
nucleons, and the two-nucleon scattering wave function $\ket{\psi_f}$.

The treatment of the nuclear current operator has been discussed in 
sect. \ref{sec2}. In the final state wave function  $\ket{\psi_f}$ only the 
interaction of each of the two nucleons with the residual nucleus is included. 
The effect of the mutual interaction between the two outgoing nucleons 
 ($NN$-FSI) has been studied in \cite{ScB03,ScB04} within a
perturbative  approach. The contribution  of $NN$-FSI depends on the 
kinematics and on the type of electromagnetic probe, 
and in particular situations produces a significant enhancement of the 
calculated cross section \cite{ScB03,ScB04}. Work is in progress to include 
this contribution within a more accurate treatment.
 Detailed studies not outlined here have shown now that $NN$-FSI do not
 disturb qualitatively the conclusions concerning the
$\Delta$-current drawn below. Consequently,  $NN$-FSI 
 can be safely  neglected in this work.
 Therefore, the scattering 
state is written as the product of two uncoupled single-particle distorted 
wave functions, eigenfunctions of a complex phenomenological spin and energy 
dependent optical potential \cite{Nad81}.

The TOF $\ket{\psi_i}$ contains information on nuclear structure and
correlations and in principle requires a calculation of the two-hole spectral
function including consistently SRC as well as long-range correlations (LRC), 
main\-ly due to collective excitations of nucleons at the nuclear surface. 
 It is well known from previous work that the cross sections are generally
sensitive  to the treatment of correlations. 
Different approaches are used in  \cite{GiP97,GPA98,BGPD04,Kadrev}.
Since the main aim of the present work is to study the uncertainties in the 
treatment of the $\Delta$-current in combination with the effect of
correlations, it can be interesting to compare results
obtained with different TOF's. Therefore, the present calculations have been 
performed with the three different approaches used in Refs. \cite{GiP97},
\cite{GPA98}, and \cite{BGPD04}.

In the simpler approach of \cite{GiP97} the TOF is given by the product of a 
coupled and antisymmetrized shell model pair function and  
a Jastrow-type central and state independent correlation function taken 
from \cite{GD}. In this approach (SM-SRC) only SRC are considered and the final
state of the residual nucleus is a pure two-hole state. For
instance, the ground state of $^{14}$C is a ($p_{1/2}$)$^{-2}$ hole in $^{16}$O.

In the more sophisticated approaches of \cite{GPA98} and \cite{BGPD04}  
the TOF's are 
obtained from the two-proton spectral function 
of $^{16}$O with a two-step procedure which includes both SRC and LRC. 
In the first step, LRC are calculated in a shell-model space large enough 
to account for the main 
collective features of the pair removal amplitude. The 
single-particle propagators used for this dressed random phase approximation 
(RPA) description of the two-particle propagator include the effect of 
both LRC and SRC. In the second step, that part of the pair removal amplitudes 
which describes the relative motion of the pair is supplemented by defect 
functions, which contain SRC, obtained by solving the Bethe-Goldstone 
equation with a Pauli 
operator which considers only configurations outside the model space where 
LRC are calculated. Different defect functions are obtained for different 
relative states using different realistic $NN$-potentials. 
In the approach of \cite{GPA98,SF} (SF-A), the non-locality of the Pauli 
operator is neglected, resulting in a set of only few defect functions, which
are essentially independent of the center-of-mass (CM) motion of the pair. 
In the more recent approach of \cite{BGPD04} (SF-B) the Pauli operator is 
computed exactly, resulting in a larger number of defect functions which have 
a more complicate state dependence. Moreover, in \cite{BGPD04} the evaluation 
of nuclear structure effects related to the fragmentation of the 
single-particle strength has been improved by applying a Faddeev technique to 
the description of the internal propagators in the nucleon self-energy 
\cite{badi01,badi02}.
The defect functions used in the present calculations are obtained from the 
Bonn-A $NN$-potential for the SF-A approach and from Bonn-C for SF-B. It was 
found, however, in \cite{GPA98} that defect functions from the Bonn-A and 
Bonn-C potentials do not produce significant differences. 
 
\begin{figure*}[ttt]
\centerline{
\resizebox{0.7\textwidth}{!}{
  \includegraphics{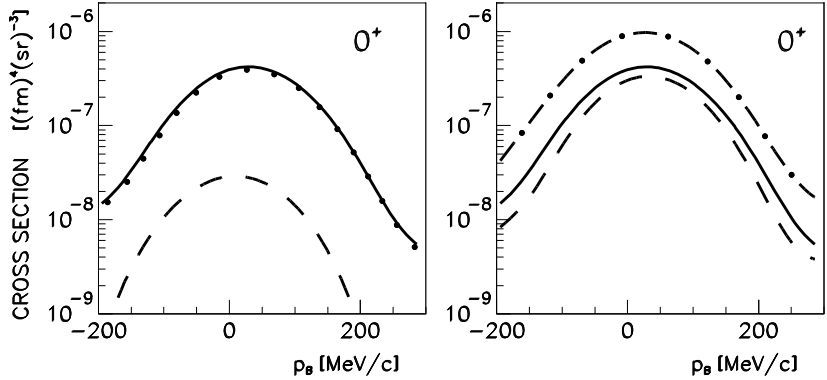}}
}
\caption{The differential cross section of the $^{16}$O$(e,e' pp)$ reaction to 
the $0^+$ ground state of $^{14}$C as a function of the recoil momentum  
$p_{\mathrm{B}}$ in a coplanar symmetrical kinematics  with $E_0= 855$ MeV, an 
electron scattering angle $\theta_e = 18^{\circ}$, $\omega=215$ MeV, and  
$k=316$ MeV/$c$. Different values of $p_{\mathrm{B}}$ are obtained changing 
the scattering angles of the two outgoing protons.  Positive (negative) values 
of $p_{\mathrm B}$ refer to situations where ${\vec p}_{\mathrm B}$ is parallel 
(anti-parallel) to ${\vec k}$. 
The $\Delta(NN)$ parametrization is used in the calculations. 
In the left panel the TOF is taken from the SF-B approach and separate 
contributions of the one-body and $\Delta$-current are shown by the dotted and 
dashed line, respectively. 
The solid curve gives the final result. The results with different TOF's are 
shown in the right panel: SF-B (solid line), SF-A (dashed line), and SM-SRC
(dash-dotted line).   
}
\label{esym}
\end{figure*}

The results for the $^{16}$O$(e,e'pp)^{14}$C$_{\mathrm{g.s.}}$ reaction in the 
symmetrical kinematics are displayed in Fig. \ref{esym}.
The shape of the recoil-momentum distribution is driven by the CM orbital
angular momentum $L$ of the knocked out pair. This feature, that is fulfilled 
in a factorized PW-approach, is not spoiled by FSI \cite{GiP97,GPA98}. 
Different partial waves of relative and CM motion are included in the TOF. For
the considered transition to the $0^+$ ground state of $^{14}$C, the main
components of relative motion are: $^1S_0$, combined with $L= 0$, and $^3P_1$, 
combined with  $L= 1$. The shape of the cross sections in Fig. \ref{esym}  
clearly indicates the dominance of the $^1S_0$, $L= 0$, component. 
The separate contributions of the one-body and $\Delta$-current, displayed in 
the left panel of the figure, show that the $\Delta$-contribution is very 
small. In practice, the whole cross section is due to the one-body current, in 
particular to its longitudinal charge-term. The contribution of the convection 
and spin-terms is practically negligible. 
The results in the left panel are obtained with the TOF from the SF-B 
approach and with the {\bf $\Delta(NN)$} parametrization. Different TOF's and 
$\Delta$-parametrizations do not change the qualitative features of the
calculated cross sections. Different parametrizations affect only the 
contribution of the $\Delta$-current, but do not affect the final result, 
that is dominated by the longitudinal part of the one-body current. Thus, in 
this case the results are insensitive to the uncertainties in the treatment of 
the $\Delta$. 
Quantitative differences are produced by the three TOF's. The results displayed 
in the right panel show that different treatments of 
correlations do not change the shape but only the size of the cross section.
The SM-SRC result is $\sim$ 2-3 times larger than SF-B, that, in turn, is
between 20-40$\%$ larger than SF-A.

\begin{figure*}[ttt]
\centerline{
\resizebox{0.5\textwidth}{!}{
  \includegraphics{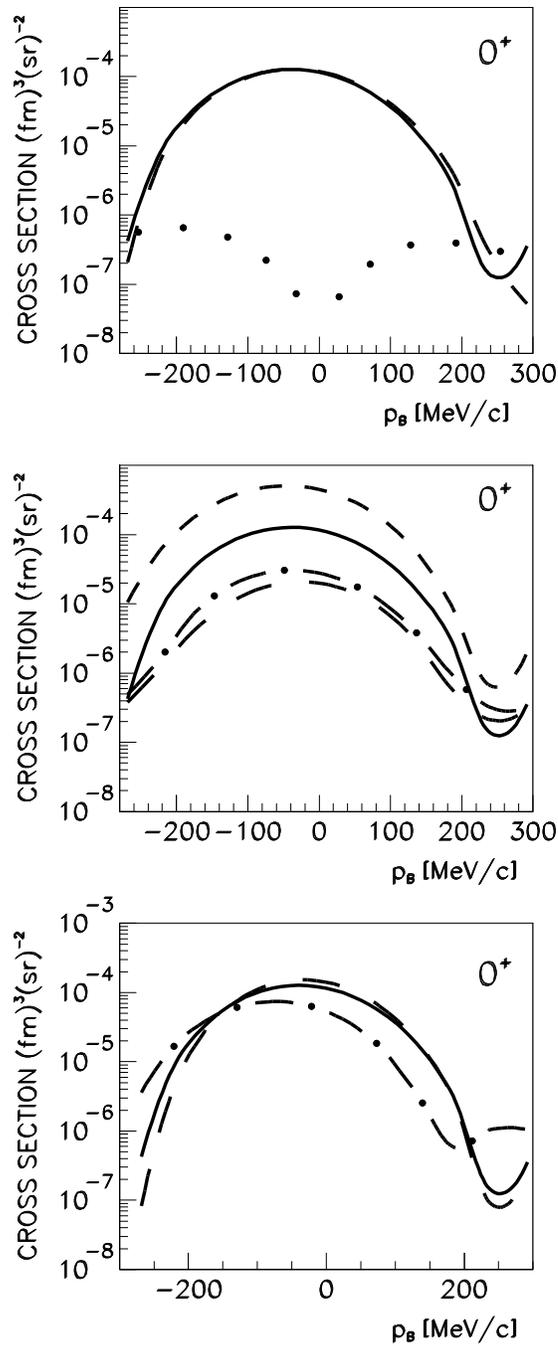}}
}
\caption{
The differential cross section of the 
$^{16}$O$(\gamma,pp)^{14}$C$_{\mathrm{g.s.}}$  reaction as a function of the 
recoil momentum  $p_{\mathrm{B}}$ in a coplanar symmetrical kinematics with 
$E_\gamma=400$ MeV. Line convention in the top and bottom panels as 
in the left and right panels, respectively, of Fig. \ref{esym}. In the middle
panel the TOF is taken from the SF-B approach and results with different 
$\Delta$-parametrizations are compared: $\Delta(NN)$ (solid line), 
$\Delta$(NoReg) (short dashed line), $\Delta(\pi N)$ (dash-dotted line), and
$\Delta (\pi N, \mathrm{mod})$ (dashed line).
}
\label{gsym}
\end{figure*}

The cross section of the $^{16}$O$(\gamma,pp)^{14}$C$_{\mathrm{g.s.}}$ reaction 
in the symmetrical kinematics at $E_\gamma = 400$ MeV, displayed in 
Fig. \ref{gsym}, is completely dominated by the $\Delta$-contribution. This
qualitative result is obtained with all the $\Delta$-par\-am\-et\-ri\-zations 
and TOF's considered in the present work. The shape of the recoil-momentum
distribution confirms also in this case the dominance of the $^1S_0$, $L= 0$, 
partial wave. It is interesting to notice, however, that the 
separate contribution of the one-body current is driven by the $L= 1$ component.
This result can be understood if we consider that in a reaction induced by a 
real photon only the transverse components of the current contribute. It has 
been demonstrated in sect. \ref{sec3} that in a symmetrical kinematics the 
transverse part of the one-body current is strongly suppressed when the two 
protons are in an initial $^1S_0$ 
relative state. Thus, the $^3P_1$, $L= 1$, component gives the main 
contribution to the one-body current. This contribution is, however, 
overwhelmed in the final cross section by the $\Delta$-current. 
Differences larger than one order of magnitude in the peak-region are 
produced by the different $\Delta$-parametrizations. The largest cross section 
is given by the $\Delta$(NoReg) prescription, but large differences are also 
found between the $\Delta(NN)$ and $\Delta(\pi N)$ results.  The modified 
propagator in $\Delta (\pi N, \mathrm{mod})$ gives only a slight reduction 
of the cross section calculated with the $\Delta(\pi N)$ parametrization.
The results produced by the TOF's from the SF-A and SF-B spectral functions 
are very close. A slight reduction and a somewhat different shape is obtained 
with the simpler SM-SRC approach. The difference in the shape 
is due to the larger contribution of the $^3P_1$, $L=1$, component in SM-SRC. 
The differences of the results with the three  TOF's in the two symmetrical 
kinematics in Figs. \ref{esym} and \ref{gsym} are due to the 
different effects of correlations on the one-body current in the 
$(e,e'pp)$ cross section, and on the two-body
$\Delta$-current in the $(\gamma,pp)$ reaction.

\begin{figure*}[ttt]
\centerline{
\resizebox{0.5\textwidth}{!}{
  \includegraphics{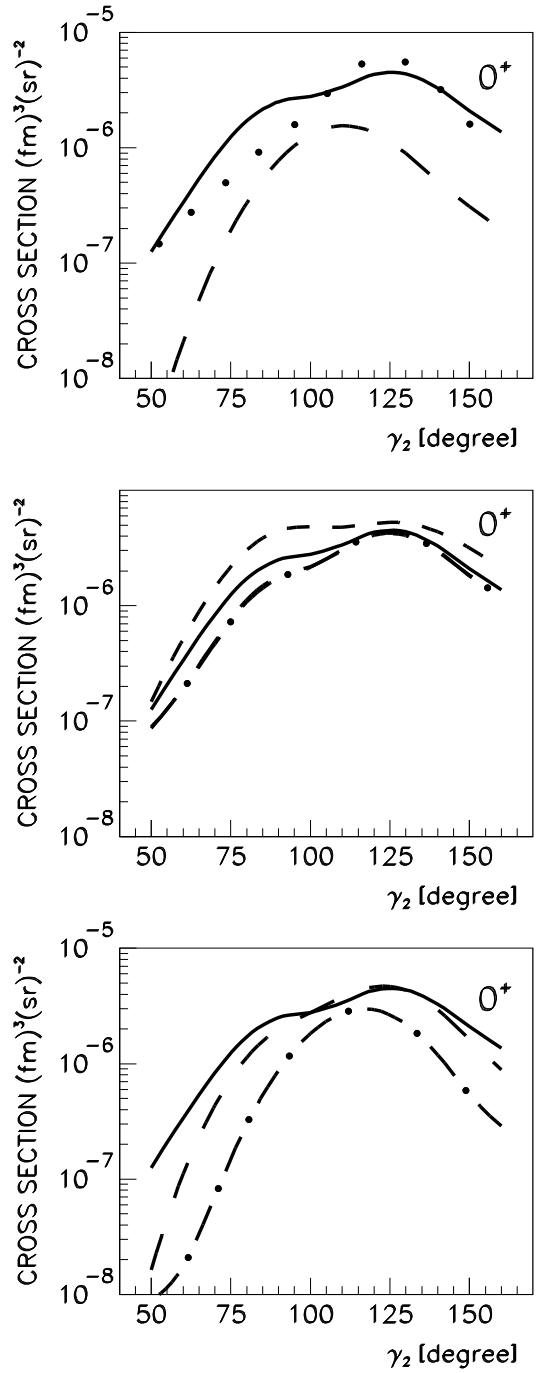}}
}
\caption{
The differential cross section of the 
$^{16}$O$(\gamma,pp)^{14}$C$_{\mathrm{g.s.}}$  reaction as a function of the 
scattering angle $\gamma_2$ of the second outgoing proton in a coplanar 
kinematics with  $E_\gamma =$ 120 MeV, $T_1 = 45$ MeV, and 
$\gamma_1=45^{\circ}$. Line convention as in Fig. \ref{gsym}.
}
\label{g120}
\end{figure*}

The results for the $^{16}$O$(\gamma,pp)^{14}$C$_{\mathrm{g.s.}}$ reaction 
in the kinematics at $E_\gamma = 120$ MeV are displayed in Fig. \ref{g120}.
In this kinematics, where  both convection- and spin-terms are important,
the one body current gives the main contribution to the cross section.  
The effect of the $\Delta$-current, however, is not negligible: it produces 
a significant enhancement of the $^3P_1$ component and a slight reduction of 
the  $^1S_0$ one. Such a reduction is due to the destructive interference 
between the spin- and the $\Delta$-current contributions in the $^1S_0$ 
relative state. The final result is that the $\Delta$ affects both the size and 
the shape of the cross section. The shape is determined by the combined effect 
of the $L=0$ and $L=1$ CM components. The uncertainties due to the various 
$\Delta$-parametrizations are within a factor of $\sim 2$. The SF-A and SF-B 
results are generally very close. The differences at lower angles are produced 
by the  $^3P_1$ component. A substantial reduction of the calculated cross 
section is obtained with the simpler SM-SRC approach.

\begin{figure*}[ttt]
\centerline{
\resizebox{0.7\textwidth}{!}{
  \includegraphics{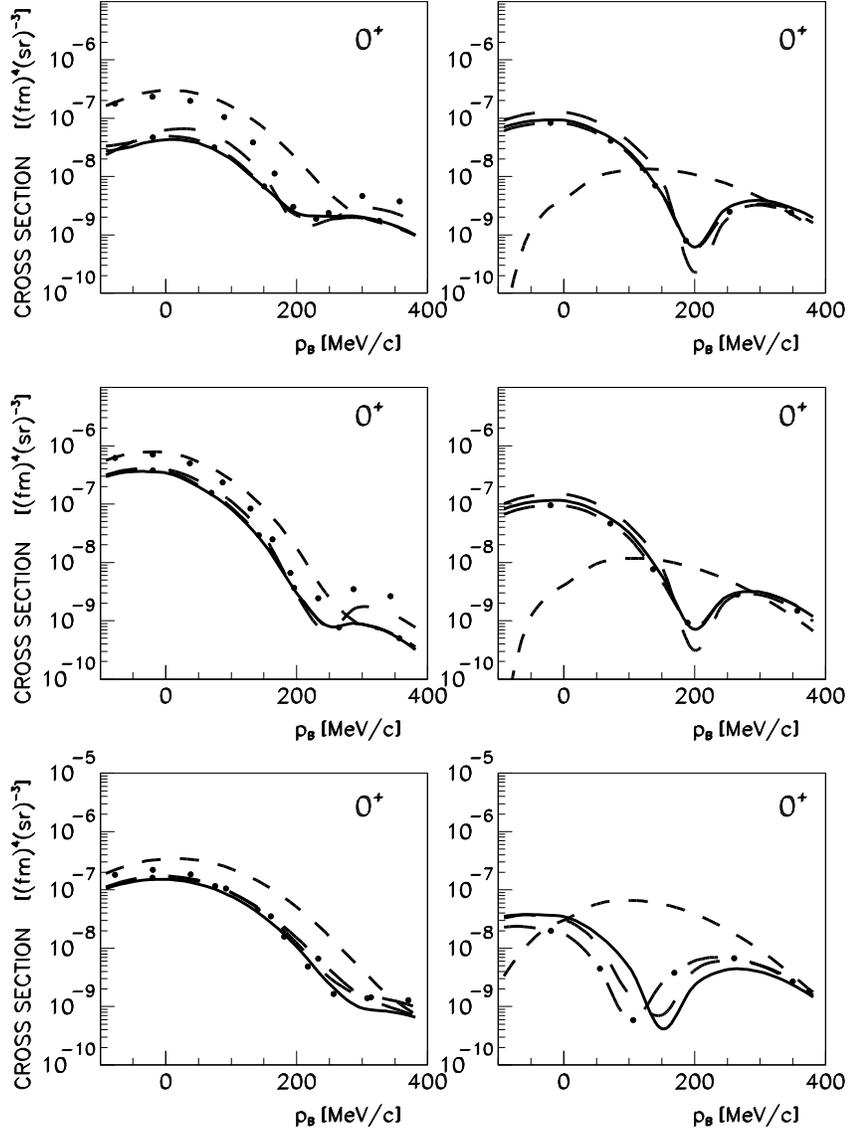}}
}
\caption{The differential cross section of the 
$^{16}$O$(e,e' pp)^{14}$C$_{\mathrm{g.s.}}$  reaction as a function of the 
recoil momentum  $p_{\mathrm{B}}$ in a super-parallel kinematics  with 
$E_0= 855$ MeV, $\theta_e = 18^{\circ}$, $\omega=215$ MeV, and  $k=316$ MeV/$c$.
Different values of $p_{\rm B}$ are obtained changing the kinetic energies of 
the outgoing nucleons. Positive (negative) values of $p_{\mathrm B}$ refer to 
situations where ${\vec p}_{\mathrm B}$ is parallel (anti-parallel) to 
${\vec k}$. Results with different TOF's are compared in the top (SF-B), middle 
(SF-A) and bottom (SM-SRC) panels. Results with different 
$\Delta$-parametrizations are displayed by solid ($\Delta(NN)$), short dashed 
($\Delta$(NoReg)), dash-dotted ($\Delta(\pi N)$), and dashed 
($\Delta (\pi N, \mathrm{mod})$) lines. The separate contributions of the 
two-body $\Delta$-current are shown in the right panels, the final results 
given by sum of the one-body and $\Delta$-currents are shown in the left 
panels. The dotted lines give the separate contribution of the one-body current.
}
\label{sp0}
\end{figure*}

The cross sections for the $^{16}$O$(e,e'pp)^{14}$C$_{\mathrm{g.s.}}$ reaction 
in the super-parallel kinematics are displayed in Fig. \ref{sp0}. The results
obtained with different TOF's and $\Delta$-par\-am\-etri\-zations are 
compared in the figure. Dramatic differences are found between the 
unregularized and the  
regularized treatments of the $\Delta$-current.  The $\Delta$-contribution  
calculated with the $\Delta$(NoReg) approach differs both in size and shape 
from the results of the three regularized versions, that are in general 
close to each other. The final cross section, given by the sum of the one-body 
and the $\Delta$-current, calculated in the $\Delta$(NoReg) approach turns 
out to be generally larger than the cross section due only to the one-body 
current. In contrast, the regularized $\Delta$-par\-am\-et\-rizations give, 
in combination with the one-body current, strong destructive interference 
effects and the final cross section is generally lower than the contribution 
of the one-body current. 
The relevance of such a destructive interference depends on the 
relative weight of the one-body and $\Delta$-current contributions, that is 
different with the different TOF's. For each TOF similar results are obtained 
with the three regularized $\Delta$-currents. The differences between the 
results of unregularized and regularized treatments in the final cross 
section depend on the TOF 
and can be large.  

The SF-A and SF-B approaches produce different one-body contributions.
A substantial reduction for low values of the recoil momentum is obtained with 
SF-B \cite{BGPD04}. In contrast, the $\Delta$-contributions obtained in the two
models are very similar. Thus, with SF-B 
the separate contributions of the one-body and $\Delta$-current are of about 
the same size, while with SF-A the one body-contribution is larger than the 
one due to the $\Delta$.  As a consequence, a stronger destructive 
interference,  as well as a larger difference between the results of the 
regularized and unregularized prescriptions, is found with SF-B.  
The uncertainties due to the  $\Delta$-treatment  are strongly reduced with 
SF-A and become even smaller with  the simpler SM-SRC approach. 
We point out the large differences obtained with the three 
TOF's both in the size and shape of the calculated cross sections. 
These differences are due to the different treatments of correlations in the 
three models, and are emphasized in this particular super-parallel kinematics 
by the interference between the one-body and the $\Delta$-current.

\begin{figure*}[ttt]
\centerline{
\resizebox{0.7\textwidth}{!}{
  \includegraphics{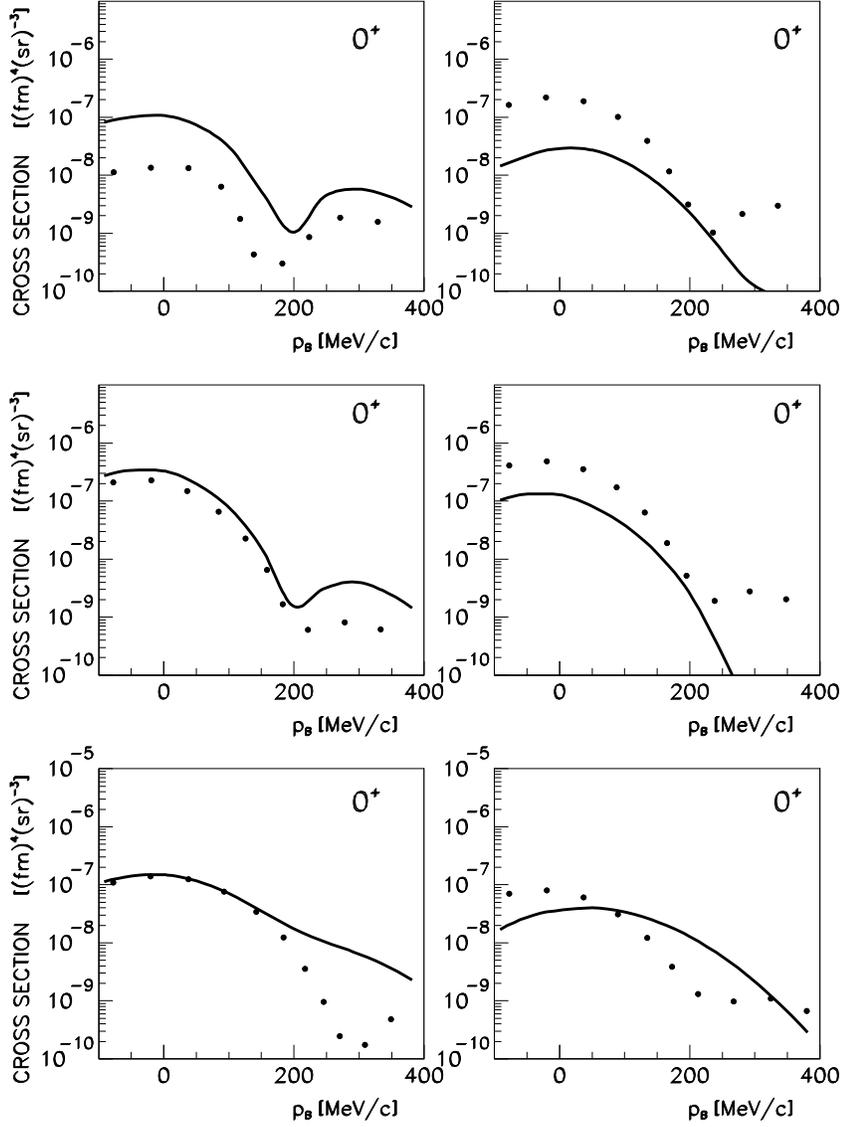}
}}
\caption{The differential cross section of the 
$^{16}$O$(e,e' pp)^{14}$C$_{\mathrm{g.s.}}$  reaction in the same kinematics as 
in Fig. \ref{sp0}. TOF's in the different panels as in Fig. \ref{sp0}. 
The dotted lines show the contribution of the one-body longitudinal charge- 
(left panels) and transverse spin-current (right panels). The solid lines give 
the results where the two-body $\Delta$-current, calculated  
with the $\Delta(NN)$ parametrization, is added to the corresponding one-body 
contribution.  
}
\label{sp00}
\end{figure*}

Some arguments concerning interference effects be\-twe\-en the $\Delta$ and the 
one-body current have been already discussed in sect. \ref{sec3}. The numerical 
results obtained in the symmetrical and super-parallel kinematics of the 
$(e,e'pp)$ reaction confirm those arguments. In order to illustrate the 
conclusions of sect. \ref{sec3} with a specific numerical example and to 
understand more thoroughly the results of Fig. \ref{sp0}, we compare in  
Fig. \ref{sp00} the separate contributions of the longitudinal charge- and 
transverse spin-currents obtained with the three TOF's. The sum of each term  
with the $\Delta$-current is also shown in the figure. The contribution of the 
convection-current is negligible and is not considered here. The calculations 
presented in the figure are performed with the $\Delta(NN)$ prescription. 
Similar results are obtained with $\Delta(\pi N)$ and 
$\Delta (\pi N, \mathrm{mod})$. 

It has been demostrated in sect. \ref{sec3} that the
contribution of the spin-term, that is minimized in symmetrical kinematics, 
becomes important in super-parallel kinematics. This result is confirmed by 
the results of Fig. \ref{sp00}. The comparison between the charge and 
spin-current contributions shows that the spin-current contribution is 
generally larger than the longitudinal charge-current one. With the SF-A 
approach the spin-current contribution turns out to be larger by one order of 
magnitude. It is interesting to notice that while only small differences are 
found between the SF-A and SF-B results for the spin-current, the 
longitudinal-current contribution calculated in the SF-B approach is about one order of magnitude 
lower than with SF-A. This explains 
the different results given by the two TOF's for the full one-body currents 
in  Fig. \ref{sp0}  and in \cite{BGPD04}. 
When added to the longitudinal term, the $\Delta$-current produces an 
enhancement of the cross section that is large with SF-B, where the 
contribution of the longitudinal term is small, and small with SF-A, where 
the longitudinal  contribution is much larger. A different effect is given by  
the $\Delta$ in combination with the spin-term. Here it produces  a strong 
destructive interference that is larger with SF-B than with SF-A. 
Similar results are found with all the regularized 
$\Delta$-parametrizations. In contrast, no destructive interference effect 
is obtained with the $\Delta$(NoReg) prescription. This explains the results 
of Fig. \ref{sp0}.  

The results displayed in  Fig. \ref{sp00} with the SF-A and SF-B approaches 
are dominated by the $^1S_0$ component, that is responsible for the 
destructive interference  between the spin- and $\Delta$-current contributions. 
For the $^3P_1$  component the $\Delta$ gives always an enhancement of the 
one-body cross section. The somewhat different results shown in Fig. \ref{sp00} 
with the SM-SRC two-nucleon wave function are due to the heavier weight of the 
$^3P_1$ component in this approach.

\begin{figure*}[ttt]
\centerline{
\resizebox{0.55\textwidth}{!}{
  \includegraphics{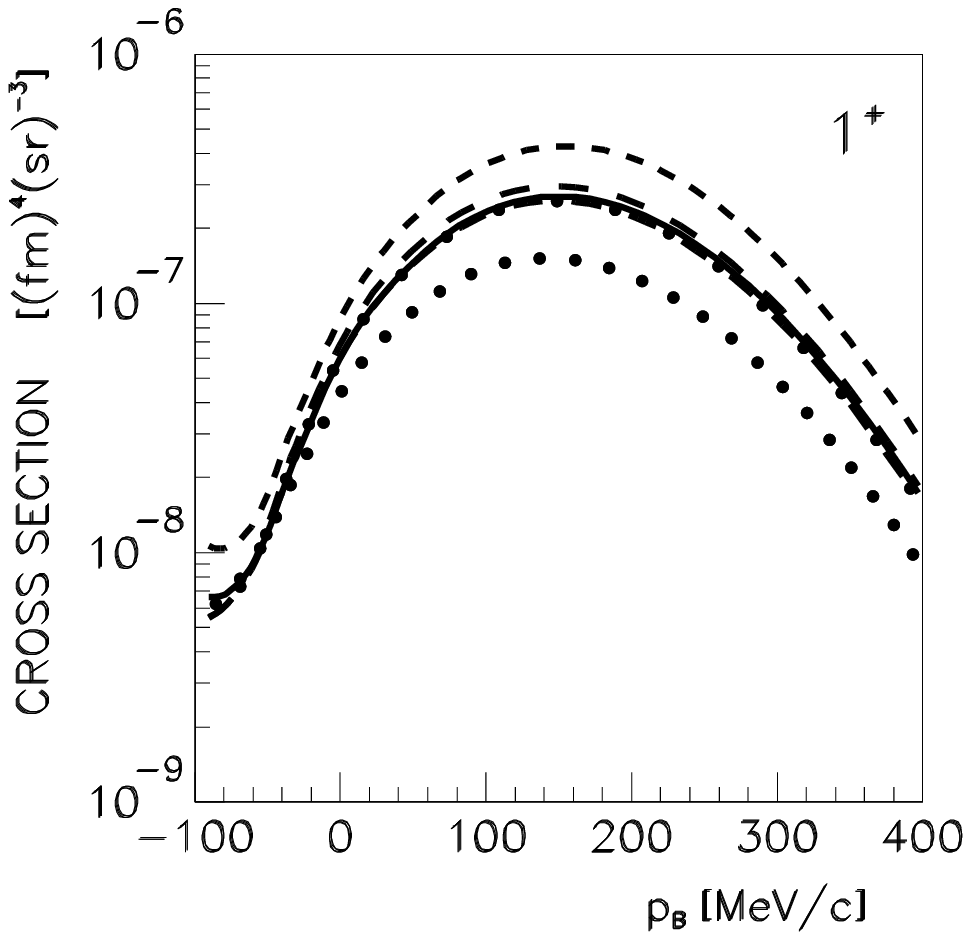}
}}
\caption{The differential cross section of the $^{16}$O$(e,e' pp)$ reaction to 
the $1^+$ excited state at 11.31 MeV of $^{14}$C in the same super-parallel 
kinematics as 
in Fig. \ref{sp0}. The TOF is taken from the SF-B approach. The dotted line 
gives the separate contribution of the one-body current, the other lines the 
final result with different $\Delta$-parametrizations: $\Delta(NN)$ 
(solid line), $\Delta$(NoReg) (short dashed line), $\Delta(\pi N)$ 
(dash-dotted line), $\Delta (\pi N, \mathrm{mod})$ (dashed line).
}
\label{sp1}
\end{figure*}

The effects of the different $\Delta$-parametrizations in the  $^3P$ states 
can be seen in Fig. \ref{sp1}, where the cross section of the 
$^{16}$O$(e,e' pp)$ reaction to the $1^+$ excited state of $^{14}$C is 
displayed in the same super-parallel kinematics as in Fig. \ref{sp0}. 
For this transition only $^3P$ states contribute: $^3P_0,^3P_1,^3P_2$, all 
combined with $L=1$.
The $\Delta$-current produces a substantial enhancement of the cross section
calculated with the one-body current. The two separate contributions are of
about the same size and  add up in the cross section. Only slight differences 
are given by the regularized $\Delta$-parametrizations. A larger cross section 
is given by the $\Delta$(NoReg) prescription. The differences, however, are
within a factor of about 2.

\section{Summary and conclusions}\label{sec5}

The combined effect of the two-body $\Delta$-current and correlations has been
discussed in electro- and photoinduced exclusive two-proton knockout reactions
from $^{16}$O. 

The $\Delta$-current operator consists of an excitation and a deexcitation 
part. The potential describing the transition $N\Delta \rightarrow NN$ via 
meson exchange contains the $\pi$- and $\rho$-exchange. Results with 
unregularized and regularized transition potentials have been compared in order to evaluate the
theoretical uncertainties in the $\Delta$-contribution. Different 
pa\-ra\-me\-tri\-zations of the effective $\Delta$-current have been proposed.
The parameters for the regularized prescriptions are fixed
alternatively from  $\pi N$- and $NN$-scat\-ter\-ing in
the  $\Delta$-region. Nuclear  medium effects have been 
included through a shift in the $\Delta$-propagator suggested by a comparison 
between inclusive electron-scattering data in the $\Delta$-region and the 
results of the $\Delta$-hole model.

Correlations are included in the two-nucleon overlap function within different 
approaches.  In a simpler treatment the overlap function is given by the 
product of a shell-model pair wave function and of a Jastrow-type correlation 
function. In a more sophisticated model the overlap function is obtained from 
the  calculation of the 
two-proton spectral function. The results of two different calculations are 
compared,  where the spectral function has been evaluated in the framework of 
a many-body approach with a realistic nuclear force, and where short-range 
and long-range correlations are taken into account consistently with a 
two-step procedure.

In the final state only the 
interaction of each of the two nucleons with the residual nucleus is included 
through an optical potential fitted to elastic proton-nucleus scattering. 
The mutual interaction between the outgoing nucleons ($NN$-FSI)
 is neglected, because it is irrelevant for the qualitative
understanding of the $\Delta$-current in the different considered kinematics. 

Many different kinematics can in principle be considered. With a few numerical 
examples we have shown that in different 
situations different reaction mechanisms can be relevant and the various 
ingredients of the calculations can affect the cross section in a different 
way. Thus, a suitable choice of kinematics can allow us to reduce the 
uncertainties on the theoretical ingredients and disentangle the 
specific contributions. 

There are situations, like the symmetrical kinematics in electron scattering, 
where the contribution of the one-body current through correlations is 
dominant, while the  $\Delta$-contribution is very small and therefore the 
cross section 
is insensitive to the uncertainites in the treatment of the $\Delta$-current.  
Such situations appear very well suited to probe correlations, even though the 
size and shape of the cross section mainly depend in this case on the momentum 
distribution of the proton pair inside the nucleus.

There are also situations, like the symmetrical kinematics in photoreactions at 
intermediate energy, where the contribution of the one-body current is 
suppressed and the cross section is dominated by the $\Delta$-current. In this
case the cross section is very sensitive to the treatment of the $\Delta$ and 
to its parameters. Such situations appear well suited to study the 
$\Delta$-current in the nucleus and can be helpful to pin down the most suitable
parametrization. 

We have shown that there are also kinematics in photoreactions at lower 
energy where the contribution of the  one-body current is competitive 
and even larger than the contribution of the $\Delta$-current. 
In this case both contributions are important, but the interference between them 
is small and the uncertainties due to different $\Delta$-par\-am\-et\-rizations are 
not large. Such situations can be helpful to investigate correlations in 
alternative or, preferably, in combination with electron scattering. 

Finally, we have considered the case of the super-par\-al\-lel kinematics in 
the $(e,e'pp)$ reaction. Here the cross section gets a large contribution from 
both one-body and two-body currents and their interference can be crucial in 
the final result. The peculiarity of the super-parallel kinematics is that an
important role is played in the one-body current by the transverse spin-term,
whose contribution is generally larger than the one due to the longitudinal
charge-term. The spin-current has a magnetic dipole form which can strongly 
interfere with the dominant term of the $\Delta$-current when the two protons
are in an initial $^1S_0$  relative state. These interference effects can be
crucial in the final cross section when the $^1S_0$ component dominates the
reaction process, like in the case of the transition to the ground state of 
$^{14}$C.
This behavior of the super-parallel kinematics is completely different from 
the symmetrical kinematics, where the spin-current is negligible and the cross
section is dominated by the charge-term.

We have found that in the super-parallel kinematics the interference between the
spin- and the $\Delta$-current in the $^1S_0$ state is always destructive when a
regularized prescription is used for the $\Delta$ and that different regularized
$\Delta$-pa\-ra\-me\-tri\-zations give similar results. 
In contrast, an unregularized current has a different behavior and gives no 
destructive interference with the spin-current. 
The difference depends mainly on the relative weight of the longitudinal
and spin-terms, that, in turn, depends on the treatment
of correlations in the overlap function. 

Different overlaps can produce large 
differences in the contribution of the longitudinal charge-current,
which is essentially an antisymmetrized amplitude given by the Fourier 
transform of the correlation function. When the charge contribution becomes 
small, the negative interference due to the spin-current makes the cross 
section small and a reduction of up to one order of magnitude can be obtained 
with respect to the one-body contribution. When the charge contribution is 
large, it is able to counterbalance the negative interference and the final 
cross section becomes larger.

We have found that suitable kinematics can be envisaged to study the different
ingredients entering the cross section and the different reaction mechanisms 
of electromagnetic two-proton knockout reactions.

Situations where the longitudinal part of the one-body current is dominant and
the uncertainties in the treatment of the $\Delta$-current are negligible are 
well suited to study short-range correlations. Peculiar interesting 
effects and a strong sensitivity to correlations are found in kinematics where 
the different terms of the nuclear current compete and interfere. If we want,
however, to extract unambiguous information on correlations, it is indispensable
that all the theoretical ingredients of the reaction are under control and, in 
particular to resolve the uncertainties in the $\Delta$-contribution. To this
purpose, situations where the $\Delta$-contribution is dominant can also be
envisaged, that are useful to study the behavior of the $\Delta$-current in the
nucleus.  

In conclusion, electromagnetic two-proton knockout reactions contain a 
wealth of
information on correlations and on the behavior of the $\Delta$-current in a
nucleus, but it seems to be impossible  to extract this interesting information
 just from one or two ``ideal'' kinematics.
Consequently, experimental data are needed in various kinematics
 which mutually supplement each other.  Concerning the different
 kinematics studied in this work, we are of course aware that a
 suitable  one  for a theoretical analysis is not necessarily
 the best  one  for an  experimental  measurement. Since
 comparison  between theory and data is necessary, close and 
continuous collaboration between theorists and experimentalists is 
 therefore essential to achieve a satisfactory understanding of 
electromagnetic two-proton knockout  and to determine all the
 ingredients  contributing to the cross section.

\centerline{\bf{ Acknowledgement}}
This work has been supported by the Deutsche Forschungsgemeinschaft
 (SFB 443) and by the Istituto Nazionale di Fisica Nucleare (INFN). M. Schwamb
 would like to thank the 
 Dipartimento di Fisica Nucleare e Teorica of the University of Pavia
 for the warm hospitality during his stays in Pavia.


\begin{thebibliography}{00}
\bibitem{BoG96} S. Boffi, C. Giusti, F.D. Pacati, M. Radici, {\it
 Electromagnetic Response of Atomic Nuclei}, in Oxford Studies in 
 Nuclear Physics (Clarendon Press, Oxford, 1996).

\bibitem{ScB03}  M. Schwamb, S. Boffi, C. Giusti, F.D. Pacati,
  Eur. Phys. J. A {\bf 17}, 7 (2003).

\bibitem{ScB04}  M. Schwamb, S. Boffi, C. Giusti, F.D. Pacati,
  Eur. Phys. J. A {\bf 20}, 233 (2004).
  
\bibitem{BGP} S. Boffi, C. Giusti, F.D. Pacati,
Phys. Rep. {\bf 266}, 1 (1993).

\bibitem{GiP97} C. Giusti, F.D. Pacati, Nucl. Phys. A {\bf 615},
373 (1997).

\bibitem{GPA98} C. Giusti, F.D. Pacati, K. Allaart, W.J.W. Geurts, W.H.
Dickhoff, H.M. M\"{u}ther, Phys. Rev. C {\bf 57}, 1691 (1998).

\bibitem{GiP98} C. Giusti, F.D. Pacati, Nucl. Phys. A {\bf 641}, 297 (1998).

\bibitem{Ryc94} M. Vanderhaeghen, L. Machenil,  J. Ryckebusch, M. Waroquier, 
Phys. Lett. B {\bf 316}, 17 (1993).

\bibitem{Ryc97} J. Ryckebusch, V. Van der Sluys, K. Heyde, H. Holvoet, W. Van
Nespen, M. Waroquier, M. Vanderhaeghen, Nucl. Phys. A {\bf 624},
581 (1997).

\bibitem{Ryc04} J. Ryckebusch, W. Van Nespen, Eur. Phys. J. A {\bf 20}, 435 
(2004).

\bibitem{Co03} M. Anguiano, G. Co', A.M. Lallena, J. Phys. G {\bf 29},
1119 (2003).
 
\bibitem{Co04} M. Anguiano, G. Co', A.M. Lallena, Nucl. Phys. A {\bf 744},
168 (2004).

\bibitem{Sch95} R. Schmidt, diploma thesis, Mainz, 1995.  

\bibitem{BM73} B.H. Brandsen and R.G. Moorhouse, {\it The Pion-Nucleus 
System} (Princeton University Press, Princeton, NJ, 1973).

\bibitem{WiA97}  P. Wilhelm, H. Arenh\"ovel, C. Giusti, F.D. Pacati,
Z. Phys. A {\bf 359}, 467 (1997).

\bibitem{Mac89} R. Machleidt, Adv. Nucl. Phys. {\bf 19}, 189 (1989).

\bibitem{GaM90} H. Garcilazo, T. Mizutani, {\it $\pi NN$ Systems} (World
Scientific, Singapore, 1990).  
 
\bibitem{VanM} M. Vanderhaeghen, L. Machenil,  J. Ryckebusch, M. Waroquier, 
Nucl. Phys. {\bf A 580}, 551 (1994).

\bibitem{Co93} J.E. Amaro, G. Co', A.M. Lallena, Ann. Phys. (N.Y.) {\bf 221},
306 (1993).

\bibitem{MaH87} R. Machleidt, K. Holinde, Ch. Elster, Phys. Rep. {\bf 149},
1 (1987).

\bibitem{Wil92} P. Wilhelm, Dissertation, Mainz, 1992.

\bibitem{ScA01} M. Schwamb, H. Arenh\"ovel, Nucl. Phys. {\bf A 690}, 
647 (2001).

\bibitem{PoS87} H. P\"opping, P.U. Sauer, X.-Z. Zang, Nucl. Phys.  A 
{\bf 474}, 557 (1987).

\bibitem{TaO85} H. Tanabe, K. Ohta, Phys. Rev. C {\bf 31}, 1876 (1985).

\bibitem{ElH88} Ch. Elster, K. Holinde, D. Sch\"utte, R. Machleidt,  
Phys. Rev.  C {\bf 38}, 1828 (1988).


\bibitem{HoM97} G. Holzwarth, R. Machleidt, Phys. Rev.  C {\bf 55},
1088  (1997). 

\bibitem{ChL88} C. R. Chen, T.-S. H. Lee, Phys. Rev. C {\bf 38},
 2187 (1988).

\bibitem{KoM84}  J.H. Koch, E.J. Moniz, N. Ohtsuka, Ann. Phys.
 (N.Y.) {\bf 154}, 99 (1984).

\bibitem{KoO85} J.H. Koch, N. Ohtsuka, Nucl. Phys. A {\bf 435}, 
 765 (1985).  

\bibitem{McG} I.J.D. MacGregor {\it et al.}, Phys. Rev. Lett.
{\bf 80}, 245 (1998).

\bibitem{BGPD04} C. Barbieri, C. Giusti, F.D. Pacati, W.H.Dickhoff, Phys. Rev. 
C {\bf 70}, 014606 (2004).

\bibitem{Rosner}  G. Rosner, Prog. Part. Nucl. Phys. {\bf 44}, 99 (2000).

\bibitem{GiP91} C. Giusti, F.D. Pacati, Nucl. Phys. A {\bf 535},
573 (1991).

\bibitem{Ond97}
C.\ J.\ G.\  Onderwater {\it et al.},
 Phys.\ Rev.\ Lett.\ \textbf{78}, 4893 (1997).

\bibitem{Ond98}
C.\ J.\ G.\  Onderwater {\it et al.},
 Phys.\ Rev.\ Lett.\ \textbf{81},  2213 (1998).

\bibitem{Sta00}
R.\ Starink {\it et al., } Phys.\ Lett. B \textbf{474}, 33 (2000).

\bibitem{McG95}
J.\ C.\ McGeorge {\it et al., } Phys.\ Rev. C \textbf{51}, 1967 (1995).

\bibitem{Lam96}
Th.\ Lamparter {\it et al., } Z.\ Phys.\ A \textbf{355}, 1 (1996).

\bibitem{McG98}
J.\ D.\ McGregor {\it et al., } Phys.\ Rev.\ Lett.\ \textbf{80}, 245 (1998).

\bibitem{Wat00}
D.\ P.\ Watts {\it et al.,  } Phys.\ Rev.\ C \textbf{62}, 014616 (2000).

\bibitem{Nad81}
A.\ Nadasen   {\it et al., }  Phys.\ Rev.\ C \textbf{23}, 1023 (1981).

\bibitem{Kadrev} D.N. Kadrev, M.V. Ivanov, A.N. Antonov, C. Giusti. F.D. 
Pacati, Phys. Rev. C {\bf 68}, 014617 (2003). 

\bibitem{GD}
C.C. Gearhart, Ph.D thesis, Washington University, St. Louis (1994);\\
C.C. Gearhart, and W.H. Dickhoff, private communication.

\bibitem{SF}
W.J.W. Geurts, K. Allaart, W.H. Dickhoff, H. M\"uther,
Phys. Rev. C {\bf 53}, 2207 (1996).

\bibitem{badi01}
C. Barbieri, W.H. Dickhoff, Phys. Rev. C {\bf 63}, 034313 (2001).

\bibitem{badi02}
C. Barbieri, W.H. Dickhoff, Phys. Rev. C {\bf 65}, 064313 (2002).

\end{thebibliography}
\end{document}